\documentclass{caosp302}
\def\gtsim{\mathrel{\hbox{\rlap{\hbox{\lower4pt\hbox{$\sim$}}}\hbox{$>$}}}}
\def\ltsim{\mathrel{\hbox{\rlap{\hbox{\lower4pt\hbox{$\sim$}}}\hbox{$<$}}}}
\def\kms{\hbox{{\rm km}\,{\rm s}$^{-1}$}}
\def\bz{\hbox{$\langle B_z\rangle$}}

\usepackage{graphicx}

\articleNo{123}
\pubyear{2005}
\volume{35}
\volnumber{3}
\firstpage{1}
\received{May 1, 2005}
\accepted{August 28, 2005}

\begin{document}

%
\hauthor{G.A. Wade, E. Alecian, C. Catala, et al.}

\title{How non-magnetic are\\ "non-magnetic" Herbig Ae/Be stars?}


%
\author{
        G.A. Wade \inst{1} 
      \and 
        E. Alecian \inst{1,}\inst{2}   
      \and 
        C. Catala\inst{2}
      \and
      S. Bagnulo\inst{3}
      \and
      J.D. Landstreet\inst{4}
      \and
      J. Flood\inst{1}
      \and
      T. B\"ohm\inst{5}
      \and
      J.-C. Bouret\inst{6}
      \and
      J.-F. Donati\inst{5}
      \and
      C.P. Folsom\inst{3}
      \and
      J. Grunhut\inst{1,}\inst{7}
      \and
      J. Silvester\inst{1,}\inst{7}
       }

%
\institute{
   Department of Physics, Royal Military College of Canada, \\
   PO Box 17000, Station 'Forces', Kingston, Ontario, Canada K7K 4B4
   \and
   Obs. de Paris LESIA, 5 place Jules Janssen, 92195 Meudon Cedex, France
   \and
   Armagh Observatory, College Hill, Armagh BT61 9DG, Northern Ireland
   \and
   Department of Physics \& Astronomy, The University of Western Ontario, London, Ontario, Canada, N6A 3K7
   \and
   Obs. Midi-Pyr\'en\'ees, 14 Avenue Edouard Belin, Toulouse, France
   \and
   Laboratoire d'Astrophysique de Marseille, Traverse du Siphon - BP 8, 13376 Marseille Cedex 12, France
   \and
   Department of Physics, Queen's University, Kingston, Canada
   }

\date{March 8, 2003}

\maketitle

\begin{abstract}
Our recent discovery of magnetic fields in a small number of Herbig Ae/Be stars has required that we survey a much larger sample of stars. From our FORS1 and ESPaDOnS surveys, we have acquired about 125 observations of some $70$ stars in which no magnetic fields are detected. Using a Monte Carlo approach, we have performed statistical comparisons of the observed longitudinal fields and LSD Stokes $V$ profiles of these apparently non-magnetic stars with a variety of field models. This has allowed us to derive general upper limits on the presence of dipolar fields in the sample, and to place realistic upper limits on undetected dipole fields which may be present in individual stars. This paper briefly reports the results of the statistical modeling, as well as field upper limits for individual stars of particular interest.
\end{abstract}

%
\section{Introduction}
\label{intr}

Observations of magnetic fields in pre-main sequence Herbig Ae/Be (HAeBe) stars can serve to address several important astrophysical problems: (1) The role of magnetic fields in mediating accretion, and the validity of models which propose that HAeBe stars are simply higher-mass analogues of the T Tau stars. (2) The origin of the magnetic fields of main sequence A and B type stars. (3) The development and evolution of chemical peculiarities and chemical abundance structures in the atmospheres of A and B type stars. (4) The loss of rotational angular momentum which leads to the slow rotation observed in some main sequence A and B type stars.

Since 2004, we have been engaged in a systematic assay of the magnetic properties of bright ($m_{\rm V}\ltsim 12$) HAeBe stars using the FORS1 spectropolarimeter at the ESO-VLT (Wade et al. 2007), and the ESPaDOnS spectropolarimeter at the CFHT (Wade et al., in preparation). We have acquired about 130 Stokes $V$ (circular polarisation) spectra of over 75 HAeBe stars, with the aim of measuring the longitudinal Zeeman effect in their spectra. 

The ESPaDOnS observations are of high resolving power ($R\sim 65000$), and provide the capability to resolve the complex line profiles presented by many HAeBe stars. The longitudinal magnetic field was measured from each observation using the standard first-moment method applied to Least-Squares Deconvolved (LSD; Donati et al. 1997) profiles. The dependence of the field diagnosis on the LSD masks was explored in detail for each star, and "clean" photospheric masks were constructed by excluding lines in the spectrum that exhibited clear contamination by emission, or other significant departures from the predictions of an LTE synthetic spectrum. In most cases we found that global departures of the metallic line spectrum were relatively small, and the improvement achieved using tuned masks was minor. The magnetic field diagnosis obtained from the ESPaDOnS data is very sensitive to the projected rotational velocity $v\sin i$, and the formal uncertainties achieved consequently span a large range of precision, from very good (for most stars with $v\sin i\ltsim 80$~\kms) to essentially useless (for some stars with $v\sin i\gtsim 150$~\kms). 

The FORS1 observations, on the other hand, are of low resolving power ($R\sim 1000-1500$). Although such spectra fail to resolve the complex profiles of most lines, they are relatively insensitive to rotational broadening. Consequently, the longitudinal magnetic fields derived from FORS1 spectra (using the linear regression method developed by Bagnulo et al. 2000) provide a relatively uniform diagnosis over a large range of $v\sin i$. 

From these surveys magnetic fields have been detected in 6 stars\footnote{One magnetic star (HD 101412) discovered using FORS1, 4 magnetic stars (V380 Ori, HD 72106, HD 190073, HD 200775) discovered using ESPaDOnS, and 1 magnetic star (HD 104237) discovered previously by Donati et al. (1997). We do not discuss here results from the survey of HAeBe stars in young open clusters, briefly introduced by Alecian et al. (these proceedings).}. As the detailed properties of these magnetic HAeBe stars are discussed by Alecian et al. (these proceedings) and Folsom et al. (these proceedings), we will only  review the general results here. The two stars modeled in great detail (HD 200775, HD 72106) show stable, oblique, dipolar magnetic fields with polar intensities of about 1~kG, low $v\sin i$, and rotation periods of several days. The two stars modeled in moderate detail (HD 190073, V380 Ori) show stable, organised magnetic fields, low $v\sin i$, and rotation periods of days, possibly years. Finally, the two stars with a small number of observations (HD 101412, HD 104237) show strong longitudinal fields and simple Stokes $V$ profiles, suggesting organised magnetic fields. Both of these stars have low $v\sin i$.

Based on these results, it is clear that some Herbig Ae/Be stars host strong, organised magnetic fields, qualitatively identical to those of the main sequence Ap/Bp stars. Here, we turn our eye to our much larger collection of $\sim 125$ observations of $\sim 70$ undetected HAeBe stars, in order to examine the extent to which our observations can constrain their magnetic properties. In other words, how non-magnetic are "non-magnetic" HAeBe stars?

\begin{figure}[t]
\centerline{\includegraphics[width=8cm,angle=-90]{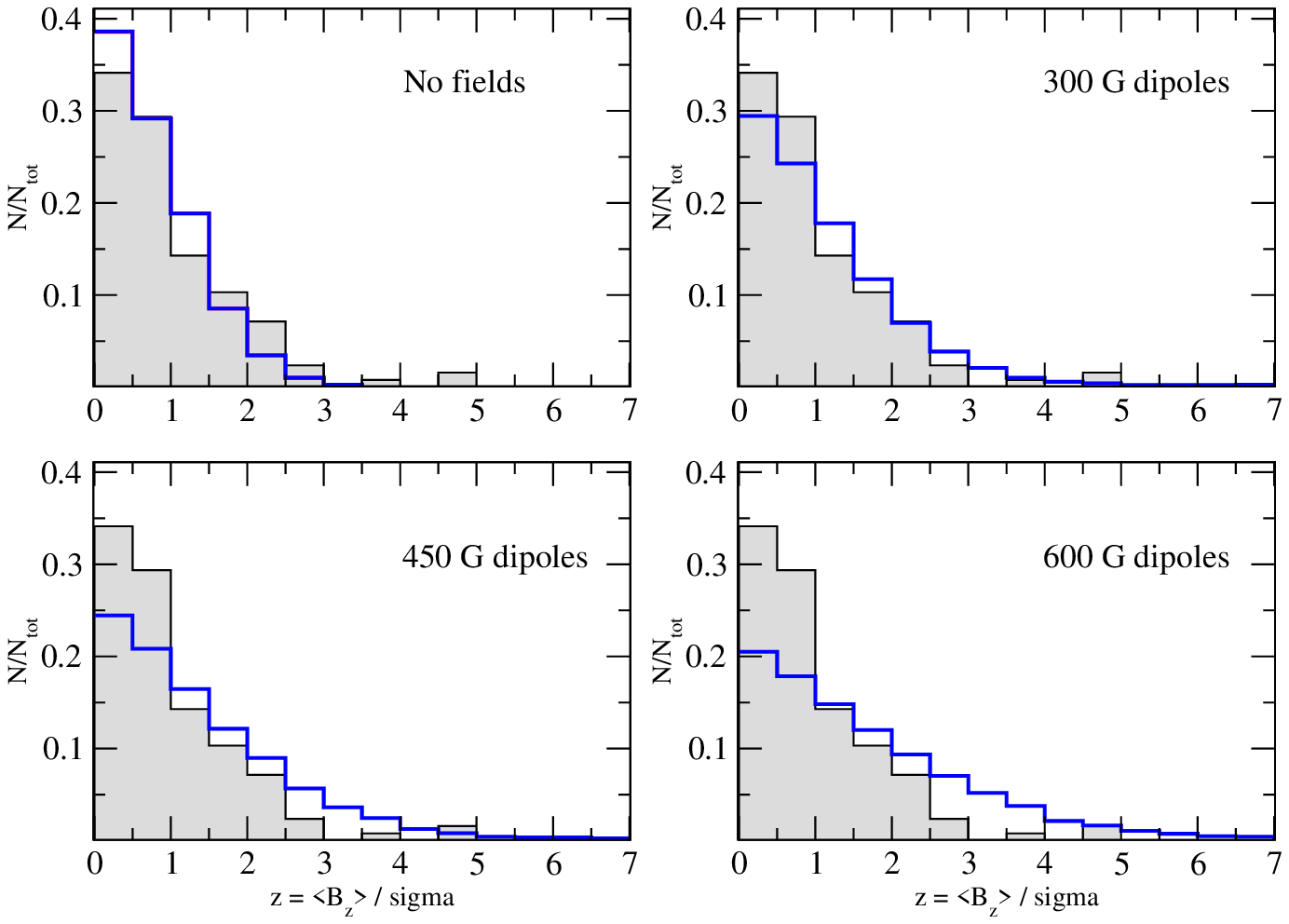}}
\caption{Observed histogram (grey filled distribution) of uncertainly-normalised longitudinal field measurements $z=\bz/\sigma_B$ of undetected HAeBe stars (including observations from FORS1 and ESPaDOnS), compared with synthetic histograms (blue unfilled distributions) corresponding to uniform populations of magnetic stars with dipole strengths as indicated.}
\label{f1}
\end{figure}

\begin{figure}
\centerline{\includegraphics[width=5cm,angle=-90]{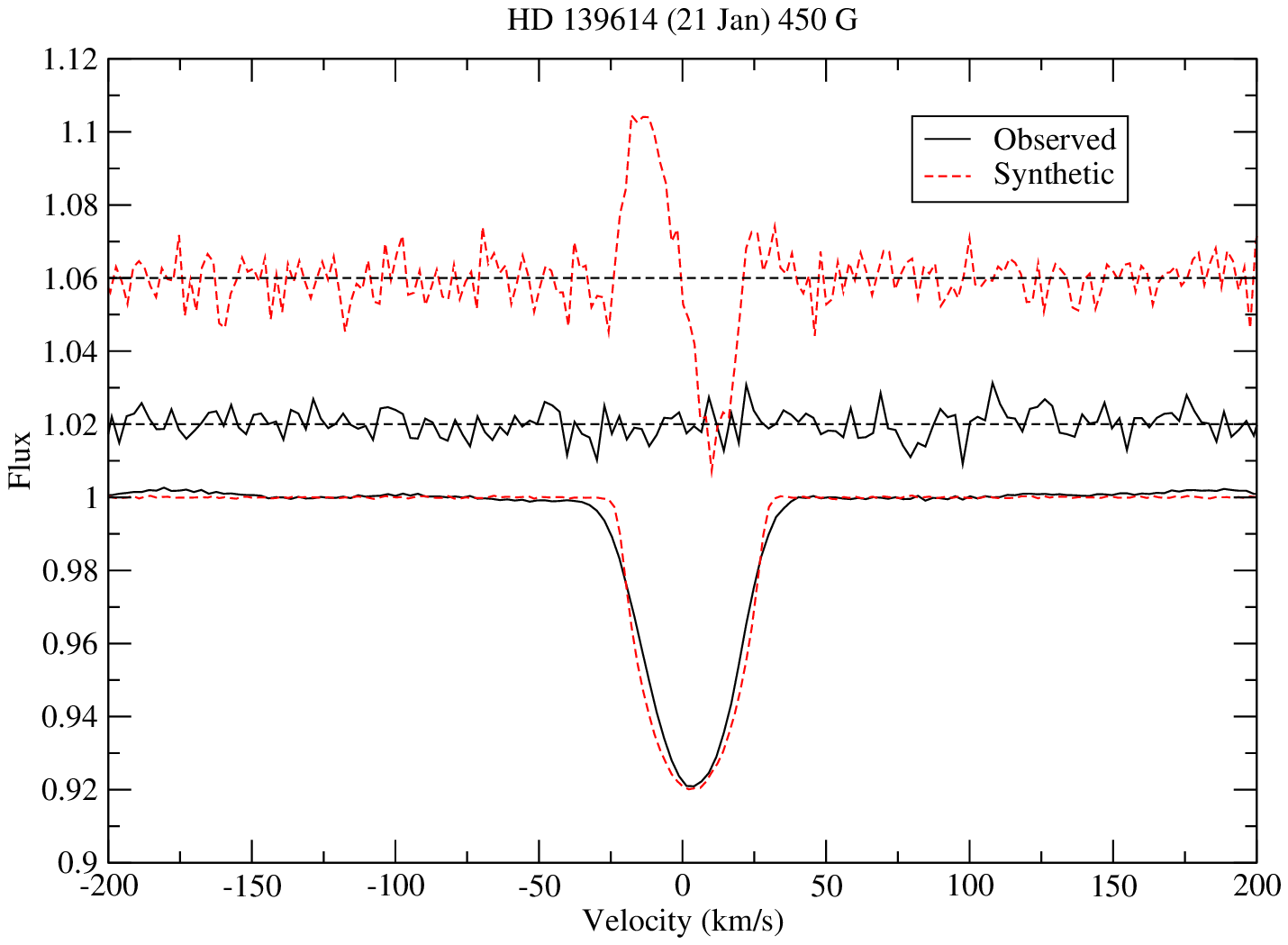}\includegraphics[width=5cm,angle=-90]{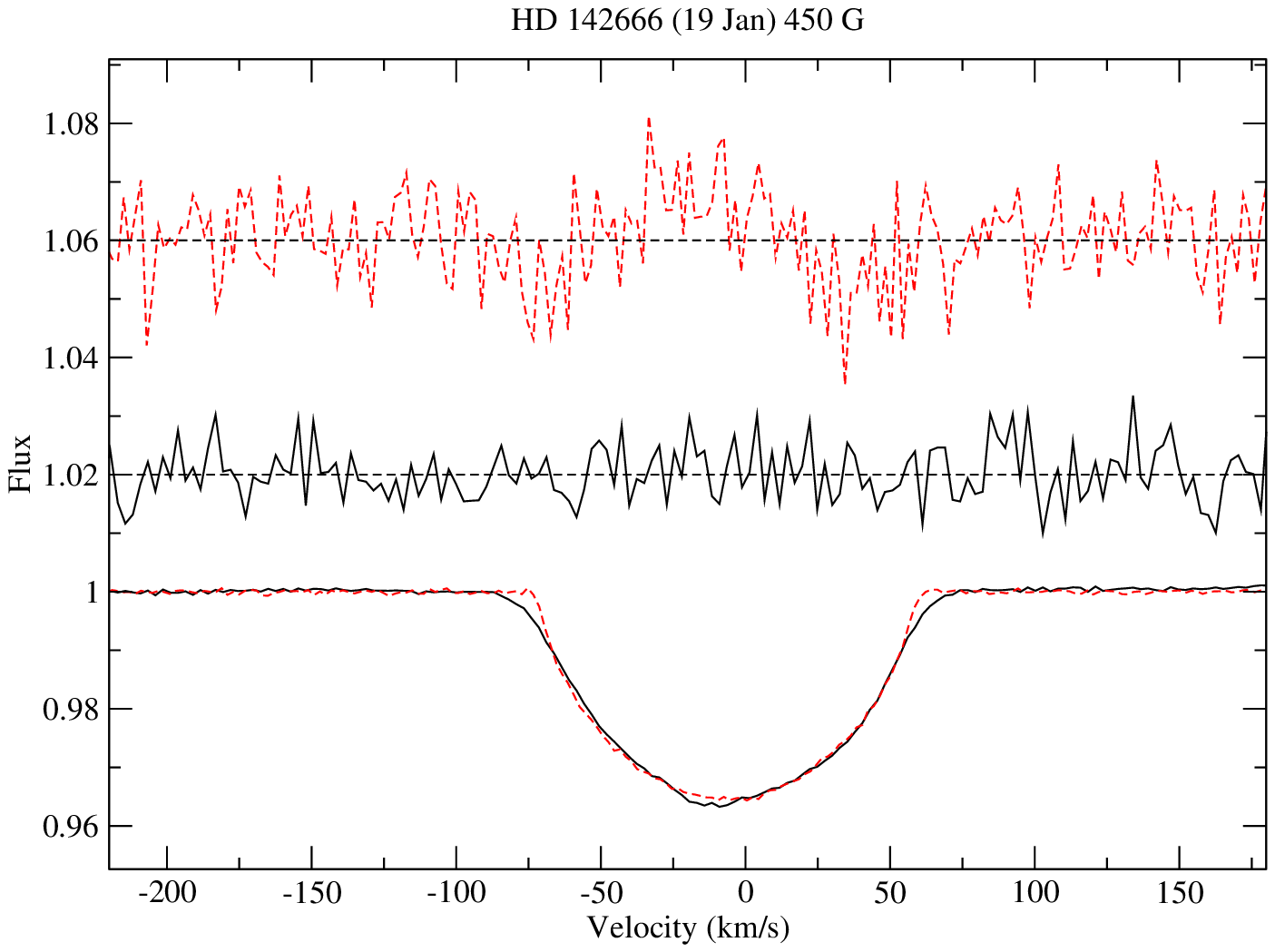}}
\caption{Observed and synthetic LSD profiles for the HAeBe stars HD 139614 and HD 142666. Solid (black) curves represent the observations, while dotted (red) lines correspond to the synthetic Stokes $I$ and $V$ profiles for a 450~G dipole field. Both simulations correspond to definite detections, indicating that such fields, if present, would be easily detected in these observations.}
\label{f1}
\end{figure}

\section{Modeling and Results}

To explore the properties of the undetected sample of HAeBe stars, we have followed a Monte Carlo approach similar to that employed by Wade et al. (2007). We developed synthetic populations of magnetic stars where each star was characterised by the inclination of its rotation axis $i$, the obliquity of its dipolar magnetic field $\beta$, the rotation phase at which it was observed $\phi$, and the intensity of its dipolar magnetic field at the magnetic pole, $B_{\rm d}$. For the purposes of our simultations, the parameters $i$ and $\phi$ were selected randomly for each star ($i$ with a $\sin i$ PDF, $\phi$ with a uniform PDF), $\beta$ was randomly set equal to either $0\degr$ or $90\degr$ (with equal probability), while the magnetic intensity $B_{\rm d}$ was fixed for all stars in a given population (and therefore defined the characteristic field strength of that population).  We then created synthetic distributions of longitudinal field measurements from each of these populations, assuming the same uncertainties characterising the real observations of undetected HAeBe stars. This procedure was repeated 100 times, using different realisations of the randomly-selected variables.

Fig. 1 compares the observed histogram of uncertainty-normalised longitudinal field measurements $z=\langle B_z\rangle/\sigma_B$ (including both FORS1 and ESPaDOnS data), with synthetic histograms compiled from the Monte Carlo simultations.

To quantitatively test whether the observed and computed distributions are representative of the same population (with the practical goal of testing if the observations imply fields which are weaker than those which characterise the models), we have performed a one-sided Kolmogorov-Smirnov (K-S) test (e.g. Conover 1971) on the cumulative distributions of longitudinal fields. The test statistic $D$ used in the K-S test is the maximum fractional difference betwen two cumulative distributions (i.e. the observed distribution and that compiled from a model). In this case, we find differences $D=0.058, 0.143, 0.228$ and 0.297 for models corresponding to dipole fields of polar intensity 0, 300, 450 and 600~G, respectively. For a sample size $N\sim 125$, a model distribution can be rejected at the 99\% level if $D\geq 0.135$. Therefore the longitudinal field measurements allow us to rule out uniform populations of stars with fields above about 300~G. On the other hand, the observations are consistent with a uniform population of stars with fields of about 300~G or smaller\footnote{We underscore that these field intensities refer to the dipole field polar strength at the stellar surface, and not to the mean longitudinal field. }.

\begin{table}[t]
\small
\begin{center}
\caption{Results of LSD profile modeling. See text for details.}
\label{t1}
\begin{minipage}{0.3\linewidth}
\begin{tabular}{rr}
\hline\hline
Model &  \#\\
   $B_{\rm d}$ & det\\
\hline
100 G & 1 \\
300 G & 2 \\
450 G & 6 \\
600 G & 6 \\
1000 G & 6 \\
2000 G& 13 \\
\hline\hline
\end{tabular}
\end{minipage}
\hspace{0.2cm}
\begin{minipage}{0.65\linewidth}
\begin{tabular}{ccrrcr}
\hline\hline
Target & Spec & $B_{\rm d}^{\rm max}$& P(\%) & \#& $\sigma_B$\\
             & Type  & (G)               &            & obs & (G)\\
\hline
HD 17081 & B8 & 100 & 95 & 2 & 9\\
HD 139614 & A6 & 300 & 98 & 3 & 14\\
HD 36112 & A4 & 450 & 96 & 4 & 30 \\
HD 142666 & A6 & 450 & 98 & 6 & 35 \\
HD 169142 & A8 & 600 & 92 & 3 & 24 \\
HD 31648 & A3 & 2000 & 95 & 2 & 52\\
HD 144432 & A9 & 2000 & 93 & 2 & 30\\
BF Ori & A5 & 2000 & 86 & 1 & 32 \\
\hline\hline
\end{tabular}
\end{minipage}

\end{center}
\end{table}

We then set out to provide a clearer evaluation of the upper limits on dipole fields for individual stars. Unfortunately, such upper limits are nearly impossible to derive using small numbers of longitudinal field measurements because the longitudinal field for most stars becomes null at some point during the rotation cycle, even in the case of a strong surface field. We have therefore employed the individual LSD profiles obtained for those stars observed using ESPaDOnS. The velocity-resolved LSD profiles allow the detection of the magnetic field even when the longitudinal field is null, thanks to the spectral separation of polarised contributions from different parts of the stellar disc due to rotational Doppler effect. However, the interpretation of an LSD profile is more complicated than a longitudinal field measurement, requiring that we create synthetic LSD profiles corresponding to each observation (reproducing its associated LSD profile depth, $v\sin i$ and signal-to-noise ratio).

To model the LSD profiles, we first fit each LSD Stokes $I$ profile with a rotationally-broadened model, to determine $v\sin i$, line depth and radial velocity. We then used the model populations of magnetic stars to create synthetic Stokes V LSD profiles corresponding to each of our observations (using the profile synthesis procedure described by Alecian et al. 2007), and introduced synthetic Gaussian noise corresponding to the noise level in the real LSD $V$ profile. Finally, for each synthetic LSD profile we evaluated the probability that a Stokes $V$ signature was detected, using the same criteria that are applied in the real LSD procedure (see Donati et al. 1997). Again, this procedure was repeated 100 times for each observation and for each model, using different realisations of the randomly-selected variables. Examples of observed and synthetic LSD profiles are shown in Fig. 2.

The results of this procedure were twofold: first, a global comparison of the predictions of each of the population models with the observations, and secondly a quantitative evaluation of the compatibility of {\em each} LSD profile with the predictions of dipole surface field models of different intensities. 

Table 1 summarises the results of this analysis. On the left, we show the number of detections of individual stars we would expect in the case of each population. Even for models as weak as 300 G, we obtain detections of small numbers of stars in over 90\% of model realisations. This result is consistent with that derived from the longitudinal field measurements, and demonstrates that the LSD profiles strongly constrain models which propose the presence of weak, organised magnetic fields in all HAeBe stars.  On the other hand, there may still exist a small number of magnetic stars, with magnetic properties similar to the detected magnetic HAeBe stars, present in our "non-magnetic" sample (but which remain undetected with the current observations). We note however that the dipole intensities derived by Alecian et al. and Folsom et al. (these proceedings) for the detected magnetic HAeBe stars, when evolved to the main sequence (Table 1 of Alecian et al.), are typical of those of the majority of Ap/Bp stars (Power et al.,  these proceedings). This indicates that the Herbig stars in which fields have been detected do not have unusually strong fields, and therefore that most of the magnetic stars in the sample are probably already detected.


On the right-hand side of Table 1, we show the inferred upper limits for dipole fields $B_{\rm d}^{\rm max}$  in individual sample stars obtained from fitting the LSD profiles. We also report the fraction of model realisations which generate a detection P(\%) when $B_{\rm d}=B_{\rm d}^{\rm max}$, the number of observations, and the derived longitudinal field error bar $\sigma_B$. Of particular interest are the lack of detections for stars in which marginal magnetic field detections had previously been claimed (HD 139614, HD 144432, HD 31648, HD 36112 and BF Ori; Hubrig et al. 2004, 2006a; Wade et al. 2007) based on observations obtained with FORS1.


\acknowledgements
EA is funded by the Marie-Curie FP6 programme, while JDL and GAW acknowledge support from NSERC and DND-ARP (Canada).


\begin{thebibliography}{}
%
%
%
%
\bibitem{} Alecian E., Catala C., Wade G.A., Donati J.-F., et al., 2007, MNRAS, in press
\bibitem{} Bagnulo S., Landstreet J. D., Lo Curto G., Szeifert T., Wade G. A., 2003, A\&A 403, 635
\bibitem{} Conover W.J., 1971, {\em Practical nonparametric statistics}, 1st ed., p. 400, Wiley (New York)
\bibitem{} Donati, J.-F., Semel, M., Carter, B. D., Rees, D. E., Cameron, A. C., 1997, MNRAS 291, 658
\bibitem{} Hubrig, S., Sch$\ddot{\rm{o}}$ller, M., Yudin, R. V., 2004, A\&A 428, L1
\bibitem{} Hubrig, S.; Yudin, R. V.; Schšller, M.; Pogodin, M. A., 2006, A\&A 446, 1089 
\bibitem{} Wade G.A., Bagnulo S., Drouin D., Landstreet J.D. and Monin D., 2007, MNRAS 376, 1145
%
%
%
\end{thebibliography}
\end{document}